\pdfoutput=1

\documentclass[11pt]{article}

\usepackage[final]{nlp4MusA}

\usepackage{times}
\usepackage{latexsym}
\usepackage{booktabs}
\usepackage{amsmath}
\usepackage[ruled,vlined]{algorithm2e}

\usepackage[T1]{fontenc}

\usepackage[utf8]{inputenc}

\usepackage{microtype}

\usepackage{inconsolata}

\usepackage{graphicx}

\title{How Far Can Pretrained LLMs Go in Symbolic Music? \\ 
Controlled Comparisons of Supervised and Preference-based Adaptation}

\author{
    \textbf{Deepak Kumar\textsuperscript{$^*$}},
    \textbf{Emmanouil Karystinaios\textsuperscript{$^*$}},\\
    \textbf{Gerhard Widmer},
    \textbf{Markus Schedl} \\
  Johannes Kepler University Linz, Austria \\
  \texttt{deepak.kumar,emmanouil.karystinaios,gerhard.widmer,markus.schedl@jku.at} }

\begin{document}
\maketitle
\def\thefootnote{*}\footnotetext{These authors contributed equally to this work}
\begin{abstract}
Music often shares notable parallels with language, motivating the use of pretrained large language models (LLMs) for symbolic music understanding and generation. Despite growing interest, the practical effectiveness of adapting instruction-tuned LLMs to symbolic music remains insufficiently characterized. We present a controlled comparative study of finetuning strategies for ABC-based generation and understanding, comparing an off-the-shelf instruction-tuned backbone to domain-adapted variants and a music-specialized LLM baseline. 
Across multiple symbolic music corpora and evaluation signals, we provide some insights into adaptation choices for symbolic music applications. We highlight the domain adaptation vs.~preserving prior information tradeoff as well as the distinct behaviour of metrics used to measure the domain adaptation for symbolic music.
\end{abstract}

\section{Introduction}

Symbolic music can be represented as discrete token sequences that exhibit long-range dependencies, hierarchical organization, and strong syntactic and structural constraints. These properties make pretrained large language models (LLMs) a great candidate for symbolic music generation and for text-conditioned musical reasoning tasks. At the same time, symbolic music differs from standard text domains in that small local errors can invalidate entire sequences, and musically meaningful structure often spans many measures, creating adaptation and evaluation challenges that are not well understood.

Recent work has demonstrated that transformer-based models can generate symbolic music and answer music-theoretic questions, either through training from scratch, music-domain pretraining, or instruction-tuned pipelines~\cite{yuan2024chatmusician,wang2025notagen,xu2024generating,bhandari2025text2midi,mundada2025wildscore}.
Of these works, ChatMusician~\cite{yuan2024chatmusician} is the only work to adapt a language model for the symbolic music domain in the literature.  
The effectiveness of adapting general LLMs to symbolic music therefore remains insufficiently characterized. In practice, researchers and practitioners face a pragmatic question: given an off-the-shelf LLM, can it be adapted to the symbolic music domain via standard domain adaptation practices of NLP as seen for other domains (e.g., medicine, law, coding)~\cite{ling2025domain,li2024fundamental,chen2025benchmarking}.  

In this work, we study the adaptation of a language model to symbolic music in a controlled setting using ABC notation, a plain-text format that encodes notes, rhythms, meter, and key as ASCII characters. Its compact structure and broad public corpus availability simplify tokenization and training compared with the more verbose **kern or XML-based encodings. We compare a base instruction-tuned backbone to domain-adapted variants trained with supervised instruction tuning and preference-based optimization, and we include a music-specialized LLM baseline for reference. To support both symbolic generation and grounded musical reasoning, we curate and unify instruction data from multiple sources and task formats, combining composition-oriented corpora with concept-focused understanding prompts paired with ABC excerpts.

Our study addresses two research questions: i) how effectively do common finetuning strategies adapt an LLM for symbolic music understanding and generation, and ii) what tradeoffs arise between music-domain gains and retention of general-domain capability. We provide empirical comparisons across symbolic corpora and general benchmarks, and we discuss vocabulary expansion through music-specific tokens as a complementary direction for improving representation and generation fidelity.




\section{Related Work}

LLMs for symbolic music have recently been explored by representing music notation as text-like sequences, enabling next-token prediction, prompting, and instruction following over symbolic scores. Music-specialized models and pipelines have demonstrated promising capabilities for composition and music-theoretic question answering when trained or adapted on large corpora of symbolic data, often using ABC, **kern, or MusicXML-derived encodings, such as ChatMusician~\cite{yuan2024chatmusician}, and MuPT~\cite{qumupt}.

Instruction tuning for music tasks extends these foundations by pairing natural-language prompts with symbolic outputs or analyses, with the goal of controllable generation and grounded reasoning about musical structure. Prior systems show that careful prompt templating and multi-stage generation can improve controllability, for example, by separating planning from realization or by providing intermediate constraints such as chord progressions or form descriptions, such as ComposerX~\cite{deng2024composerx}, MuseCoco~\cite{lu2023musecoco}, etc.

Training from scratch on symbolic music, typically with music-native objectives and large-scale corpora, remains a strong alternative to adapting general LLMs. Such models can benefit from domain-specific inductive bias in data and tokenization, and may achieve strong distributional fidelity and coherence when the training scale is sufficient, e.g. NotaGen~\cite{wang2025notagen}.

Alignment and preference learning methods, widely used in NLP to refine instruction-following behavior, have also begun to appear in music settings~\cite{wang2025notagen}. Techniques such as preference-based optimization and related alignment objectives can be used to bias generation toward outputs that satisfy structural constraints, stylistic goals, or user intent, potentially reducing the gap between likelihood-based training and human-facing quality criteria.

{
\begin{algorithm}[t]
\scriptsize
\caption{ABC Degradation Pipeline}
\label{alg:corrupt_abc}
\KwIn{ABC notation string $A$, maximum removable bars $B_{\max}$}
\KwOut{Degraded ABC notation string $A'$}

Split $A$ into a list of lines $L$\;
Initialize $L' \leftarrow L$\;

\BlankLine
\textbf{Key Changing}\;
\ForEach{line $l \in L'$}{
    \If{$l$ is a key declaration}{
        Replace key signature with a fixed alternative key (e.g., D\#)\;
    }
}

\BlankLine
\textbf{Random Pitch Swap}\;
Define note pool $P = \{C, D, E, F, G, A, B, c, d, e\}$\;
\ForEach{line $l \in L'$}{
    \If{$l$ is not a metadata line}{
        \ForEach{character $c$ in $l$}{
            With probability $p$, replace $c$ with a random note from $P$ if $c \in P$\;
        }
    }
}

\BlankLine
\textbf{Truncate bars}\;
Extract all musical (non-metadata) lines $M$ from $L'$\;
\If{$|M| > 1$}{
    Sample $k \sim \mathcal{U}(1, \min(B_{\max}, |M| - 1))$\;
    Randomly select $k$ indices from $M$\;
    Remove the selected lines from $M$\;
    Set $L' \leftarrow M$\;
}

\BlankLine
\Return Concatenate lines in $L'$ to form $A'$\
\end{algorithm}}

\begin{table}[t]
\centering
\resizebox{\columnwidth}{!}{%
\begin{tabular}{l|rr|rr}
\toprule
Statistic & Short Train & Short Test & Long Train & Long Test \\
\midrule
\# Samples           & 88,575 & 22,306 & 104,270 & 26,229 \\
Avg. Input Length    & 178.2 & 178.4 & 201.7 & 199.8 \\
Avg. Target Length   & 275.8 & 275.4 & 4,808.4 & 4,733.4 \\
Max Target Length    & 500 & 500 & 615,538 & 710,425 \\
Avg. Bars            & 10.2 & 10.3 & 233.0 & 229.1 \\
Notes per Bar        & 6.11 & 6.14 & 4.79 & 4.77 \\
\bottomrule
\end{tabular}%
}
\caption{Dataset statistics for short- and long-target ABC datasets.}
\label{tab:data_stats}
\end{table}

\begin{table*}[t]
\centering
\resizebox{\textwidth}{!}{%
\small
\begin{tabular}{lccccccc cccccc}
\toprule
\textbf{Model} 
& \multicolumn{2}{c}{\textbf{Musicpile-Short}} 
& \multicolumn{2}{c}{\textbf{PDMX-Short}} 
& \multicolumn{2}{c}{\textbf{ABCTunes-Short}} 
& \multicolumn{1}{c}{\textbf{MMLU}} & \multicolumn{2}{c}{\textbf{Musicpile-Long}} 
& \multicolumn{2}{c}{\textbf{PDMX-Long}} 
& \multicolumn{2}{c}{\textbf{ABCTunes-Long}}\\
\cmidrule(lr){2-3}
\cmidrule(lr){4-5}
\cmidrule(lr){6-7}
\cmidrule(lr){9-10}
\cmidrule(lr){11-12}
\cmidrule(lr){13-14}
& PPL$\downarrow$ & FMD $\downarrow$ 
& PPL$\downarrow$ & FMD $\downarrow$
& PPL$\downarrow$ & FMD $\downarrow$
&  $\uparrow$
& PPL$\downarrow$ & FMD $\downarrow$ 
& PPL$\downarrow$ & FMD $\downarrow$
& PPL$\downarrow$ & FMD $\downarrow$\\
\midrule
LLaMA 3.1 Inst. 8B & $693.9_{0.0}$ & $66.5_{0.2}$ & $9.0_{0.0}$ & $320.4_{0.1}$ & $28.2_{0.0}$ & $467.2_{0.3}$ & $0.59_{0.0}$ & $6.0_{0.0}$ & $130.1_{0.0}$ & $3.3_{0.0}$ & $475.8_{0.0}$ & $5.1_{0.0}$ & $449.6_{0.1}$ \\
~\cite{weerawardhena2025llama} & & & & & & \\

ChatMusician & $502.3_{0.0}$ & $66.0_{0.3}$ & $3.5_{0.2}$ & $298.0_{0.0}$ & $9.0_{0.0}$ & $172.0_{0.0}$ & $0.24_{0.0}$ & $3.0_{0.0}$ & $250.7_{1.3}$ & $2.3_{0.6}$ & $\text{na}$ & $2.8_{0.0}$ & $\text{na}$ \\
\midrule

\multicolumn{8}{l}{\textbf{Trained on short music sequences}} \\
\midrule

DPO & $401.5_{14.2}$ & $72.7_{0.0}$ & $9.0_{0.4}$ & $349.6_{0.0}$ & $517.3_{3.9}$ & $348.1_{0.0}$ & $0.31_{0.0}$ & $6.6_{0.2}$ & $233.1_{10.5}$ & $3.4_{0.3}$ & $502.0_{0.0  }$ & $21.3_{13.4}$ & $240.5_{0.0}$ \\
SFT & $36.3_{0.5}$ & $362.1_{1.5}$ & $3.7_{0.0}$ & $350.3_{0.6}$ & $9.3_{0.0}$ & $265.25_{23.4}$ & $0.29_{0.0}$ & $4.5_{0.3}$ & $291.3_{0.0}$ & $2.4_{0.0}$ & $455.5_{5.0}$ & $3.4_{0.0}$ & $253.9_{7.5}$ \\

\midrule
\multicolumn{8}{l}{\textbf{Trained on long music sequences}} \\
\midrule

DPO & $432.7_{0.0}$ & $72.4_{0.0}$ & $7.2_{0.0}$ & $327.1_{0.0}$ & $27.4_{0.0}$ & $356.5_{0.0}$ & $0.33_{0.0}$ & $5.7_{0.0}$ & $130.0_{2.0}$ & $3.2_{0.1}$ & $496.3_{4.2}$ & $5.7_{0.2}$ & $351.7_{19.2}$ \\
SFT & $56.2_{0.0}$ & $67.5_{0.0}$ & $3.7_{0.0}$ & $366.6_{0.0}$ & $12.7_{0.0}$ & $218.8_{0.0}$ & $0.27_{0.0}$ & $4.7_{0.0}$ & $130.9_{5.5}$ & $1.9_{0.0}$ & $433.0_{0.8}$ & $4.0_{0.3}$ & $209.2_{10.4}$ \\
\bottomrule
\end{tabular}%
}
\caption{Perplexity (PPL) and Frechét Music Distance (FMD) across short sequence length symbolic music datasets and MMLU.\\
\scriptsize{We report \textbf{"na"} for ChatMusician as results could not be calculated due to non-redable generation for FMD calcualtion and extremely slow evaluation due to lack padding, which restricted in resolving the issue} }
\label{tab:music_llm_results_short}
\end{table*}

\section{Method}

\subsection{Task and Evaluation Scope}

We focus on symbolic music in ABC notation and study two tasks: (1) generation or continuation, where the model produces ABC sequences conditioned on prompts or partial contexts, and (2) answering concept-rooted questions about music using ABC excerpts and metadata as context.
In relation to both capabilities, we focus on symbolic music data from different sources, split into short sequence lengths and long sequence lengths. This setup allows us to study how LLMs adapt to symbolic music when trained on ABC data sources of varying quality and under different context length constraints.

\subsection{Data Curation and Unification}

We build a unified instruction-style training set by combining multiple sources that cover complementary aspects of symbolic music. We include composition-oriented corpora, such as collections of ABC tunes (monophonic) and structured symbolic datasets such as PDMX, high-quality classical music translated to ABC such as the Distant Listening Corpus (DLC)~\cite{hentschel2025dlc}, Open Lieder~\cite{openlieder} and Open String Quartets~\cite{open-string-quartets}, and understanding-oriented prompts based on MusicPile-sft~\cite{yuan2024chatmusician}. For the latter, we filter out examples dominated by caption-like or metadata-centric content (for example, YouTube-style descriptions) to better emphasize grounded musical concepts, terminology, and reasoning. All sources are normalized into a common ABC-centric schema with consistent separators and explicit fields for prompt, context, and target.

The data is split into two sets: long and short, based on the average sequence length of the target across the datasets. In the unified dataset, this splitting criterion is a sequence length of 500. Note that for short sequence length, the data from only three sources qualify, while for long sequence length, data from all sources qualify.  Table~\ref{tab:data_stats} provides the key statistics of the unified datasets' splits. 


\subsection{Finetuning Strategies}

We compare several practical adaptation strategies that are common in NLP but underexplored for symbolic music. Our primary baseline is an off-the-shelf instruction-tuned LLM (LLaMA 3.1 Inst. 8B)~\cite{weerawardhena2025llama}. We then train a supervised finetuning (SFT)~\cite{ouyang2022training} variant on the unified instruction data, optimizing next-token likelihood over the concatenated prompt, context and target with standard token-level cross-entropy loss for next token prediction. In addition, we train a preference-based variant using paired outputs for the same prompt, optimizing a contrastive objective that increases preference for the chosen output relative to a rejected alternative.

For preference-based training, we construct rejected outputs via a set of musically motivated degradations applied to the chosen target, as specified in Algorithm~\ref{alg:corrupt_abc}, while preserving ABC syntactic validity throughout. This provides a controllable source of negative samples that differ from the chosen output primarily in musical quality rather than formatting errors. We treat this signal as weak preference supervision rather than direct human preference data. Then we use the direct preference optimization (DPO)~\cite{rafailov2023direct} technique to train the preference-based model.

We report results for Base, SFT, and DPO variants, and we include ChatMusician as a music-specialized reference point in our comparison. To maintain comparability with ChatMusician, we use a model from the LLaMA family~\cite{touvron2023llama}. We further adopt an instruction-tuned version, as instruction tuning has been shown to facilitate more effective knowledge acquisition under continued training~\cite{jiang2024instruction}.

\subsection{Metrics}

We want to evaluate the LLM's performance on different music datasets to show understanding (MusicPile) and generation (rest) capabilities. We use perplexity (PPL) to measure the token-level closeness of generation to the target for understanding and music generation. In contrast, Fréchet Music Distance~\cite{retkowski2025frechet} (FMD) over CLAMP2~\cite{wu2025clamp} is used to measure the global-level musical similarity of the generated music.
We also measure the change in prior capabilities of the LLM. In this regard, we use the Massive Multitask Language Understanding (MMLU) benchmark~\cite{hendrycksmeasuring} to see the cost of adapting an LLM for the music domain. MMLU is a multiple-choice question answering benchmark over 57 different subjects (e.g., history, law) with varying levels of expertise.



\section{Results and Analysis}

Table~\ref{tab:music_llm_results_short} presents the results over the short sequence length music data and the MMLU benchmark. We observe that for music understanding (i.e., Musicpile), all models show improvement compared to the base model, and the SFT variant appears to be the best-performing. For music generation, if we focus on token-level similarity (i.e., PPL), we observe ChatMusician and SFT are performing well while DPO deteriorates. In contrast, global-level similarity, as assessed using FMD, does not exhibit clear trends across datasets. Prior knowledge retention seems inversely related to global-level similarity, with DPO least affected and ChatMusician most affected. The LLM trained on long sequences also showed some improvement on the short sequence test set. However, the improvement is generally weaker than the models trained on short sequences. In contrast to the short sequence test set, Table~\ref{tab:music_llm_results_short} shows that the long sequence test set proves to be harder to adapt for all variants.


\section{Discussion and Limitations}


\paragraph{General Capability Retention.} Finetuning for symbolic music can degrade general-domain performance, particularly when training is narrow in domain or overly aggressive in optimization. This tradeoff is especially relevant for instruction-tuned backbones that are expected to maintain broad conversational and reasoning skills. In practice, preserving general capability may require careful data mixing, conservative training budgets, and multi-objective training. The inclusion of a music-specialized baseline provides a useful reference point: music-pretrained systems can deliver strong in-domain behavior, but may deteriorate in general instruction-following and out-of-domain robustness.

\paragraph{Evaluation Considerations.} 
Likelihood-based metrics such as perplexity capture token-level fit to a corpus, but they do not fully reflect musical quality, long-range structure, or instruction adherence. Distributional proxy metrics, such as FMD, can complement perplexity, but their correlation with human judgments for symbolic music is imperfect and may depend on representation and decoding settings. Overall, we highlight that metrics for symbolic music generation and understanding might be insufficient or incomplete.

\paragraph{Limitations.} 
First, our preference-based training uses weak preference signals induced by musically motivated degradations while enforcing ABC syntactic validity. This reduces formatting confounds but does not replace human preference data, and it may bias learning toward specific error patterns. Second, our study focuses on ABC as a compact text representation; results may not transfer directly to other symbolic encodings with different granularity and constraints. Third, while we curate and unify multiple data sources, coverage remains uneven across genres, instrumentation, and levels of theoretical sophistication. Fourth, evaluation of “understanding” remains sensitive to dataset construction and prompt design, and broader task coverage is left for future work. Finally, compute and context-length constraints limit exploration of very long-form structure and multi-part scores.

\section{Conclusion}

In this work, we presented a controlled comparative study of practical finetuning strategies for adapting instruction-tuned LLM to symbolic music in ABC notation, spanning generation and concept-rooted understanding. By comparing an off-the-shelf backbone, domain-adapted variants, and a music-specialized reference model, we characterized consistent tradeoffs between in-domain symbolic music behavior and retention of general language capabilities. \\
As future work, we plan to evaluate vocabulary expansion by introducing music-specific special tokens that capture recurring ABC constructs and structural markers (for example, barlines, section labels, voice separators, or compact representations of common ornaments). This requires extending the tokenizer, initializing new embeddings, and adapting the output projection head. While left for future work, we outline this procedure as a promising way to reduce fragmentation of musically meaningful units and improve generation fidelity under limited finetuning budgets.
As an additional adaptation direction, we will complete vocabulary expansion experiments, expand the task suite and evaluation toward deeper music-theoretic understanding and longer-horizon generation, and investigate complementary mechanisms such as vocabulary expansion and structured prompting or planning to improve coherence, controllability, and grounded reasoning.

\section*{Acknowledgments}

This work was supported by the European Research Council (ERC) under Horizon 2020 grant \#101019375 “Whither Music?”, the Austrian Science Fund (FWF): Cluster of Excellence \href{https://www.bilateral-ai.net/home}{\textcolor{blue}{\textit{Bilateral Artificial Intelligence}}} (\url{https://doi.org/10.55776/COE12}) and the doc.funds.connect project \href{https://dfc.hcai.at/}{\textcolor{blue}{\textit{Human-Centered Artificial Intelligence}}} (\url{https://doi.org/10.55776/DFH23}). For open access purposes, the authors have applied a CC BY public copyright license to any author-accepted manuscript version arising from this submission.

\bibliography{nlp4MusA}

\appendix



\end{document}